\documentclass[12pt,epsf]{article}
\usepackage{amsmath}

\newcommand{\alp}{\alpha}
\newcommand{\bt}{\beta}
\newcommand{\gm}{\gamma}
\newcommand{\Gm}{\Gamma}
\newcommand{\dlt}{\delta}
\newcommand{\Dlt}{\Delta}
\newcommand{\ep}{\epsilon}
\newcommand{\tht}{\theta}

\newcommand{\vtht}{\vartheta}
\newcommand{\kp}{\kappa}
\newcommand{\lmd}{\lambda}
\newcommand{\Lmd}{\Lambda}
\newcommand{\sgm}{\sigma}
\newcommand{\Sgm}{\Sigma}
\newcommand{\vph}{\varphi}
\newcommand{\omg}{\omega}
\newcommand{\dalp}{\dot{\alpha}}

\newcommand{\bsgm}{\bar{\sigma}}

\newcommand{\be}{\begin{equation}}
\newcommand{\ee}{\end{equation}}
\newcommand{\bea}{\begin{eqnarray}}
\newcommand{\eea}{\end{eqnarray}}
\newcommand{\eql}{\!\!\!&=\!\!\!&}

\newcommand{\defa}{\!\!\!&\equiv\!\!\!&}

\newcommand{\exch}{\leftrightarrow}

\newcommand{\tl}[1]{\tilde{#1}}

\newcommand{\tr}{{\rm tr}}

\newcommand{\diag}{{\rm diag}}
\newcommand{\der}{\partial}
\newcommand{\dr}{\!\!d}
\newcommand{\hc}{{\rm h.c.}}
\newcommand{\ie}{{\it i.e.}}

\newcommand{\vev}[1]{\langle #1 \rangle}

\newcommand{\brkt}[1]{\left( #1 \right)}
\newcommand{\brc}[1]{\left\{ #1 \right\}}
\newcommand{\sbk}[1]{\left[ #1 \right]}
\newcommand{\abs}[1]{\left| #1 \right|}

\renewcommand{\Im}{{\rm Im}}

\newcommand{\cA}{\phi}
\newcommand{\cD}{{\cal D}}
\newcommand{\cF}{{\cal F}}
\newcommand{\cH}{{\cal H}}

\newcommand{\cN}{{\cal N}}
\newcommand{\cO}{{\cal O}}

\newcommand{\cW}{{\cal W}}

\newcommand{\gomg}[2]{\omega_{#1}^{\;\;#2}}
\renewcommand{\ge}[2]{e_{#1}^{\;\;#2}}
\newcommand{\gf}[2]{f_{#1}^{\;\;#2}}

\newcommand{\cDh}{{\cal D}^{\rm h}}

\newcommand{\udl}[1]{\underline{#1}}

\newcommand{\phcl}{\phi^{\rm h}_{\rm bg}}
\newcommand{\bphcl}{\bar{\phi}^{\rm h}_{\rm bg}}
\newcommand{\fhcl}{f^{\rm h}_{\rm bg}}
\newcommand{\fG}{f^G_{\rm bg}}
\newcommand{\dQ}{\delta}
\newcommand{\Dkl}{D^{SO(2)}}
\newcommand{\Abg}{\cA_{\rm bg}^\Sgm}

\newcommand{\hfbg}{\hat{f}_{\rm bg}^\Sgm}
\newcommand{\uR}[1]{u_{{\rm R}#1}^i}
\newcommand{\uI}[1]{u_{{\rm I}#1}^i}
\newcommand{\Mp}{M_{\rm pl}}

\newcommand{\NP}[1]{{\it Nucl.~Phys.}~{\bf #1}}
\newcommand{\PL}[1]{{\it Phys.~Lett.}~{\bf #1}}

\newcommand{\PR}[1]{{\it Phys.~Rev.}~{\bf #1}}
\newcommand{\PRL}[1]{{\it Phys.~Rev.~Lett.}~{\bf #1}}
\newcommand{\PTP}[1]{{\it Prog.~Theor.~Phys.}~{\bf #1}}

\newcommand{\JH}[1]{{\it JHEP}~{\bf #1}}

\begin{document}

\begin{titlepage}
\null
\begin{flushright}
 {\tt hep-th/0403295}\\
KAIST-TH 2004/01
\\
March, 2004
\end{flushright}

\vskip 2cm
\begin{center}
\baselineskip 0.8cm
{\LARGE \bf Geometry mediated SUSY breaking}

\lineskip .75em
\vskip 2.5cm

\normalsize

{\large\bf Yutaka Sakamura}{\def\thefootnote{\fnsymbol{footnote}}
\footnote[1]{\it e-mail address:sakamura@muon.kaist.ac.kr}}

\vskip 1.5em

{\it Department of Physics, \\
Korea Advanced Institute of Science and Technology \\
Daejeon 305-701, Korea}

\vspace{18mm}

{\bf Abstract}\\[5mm]
{\parbox{13cm}{\hspace{5mm} \small
We investigate SUSY breaking mediated through the deformation 
of the space-time geometry due to the backreaction of 
a nontrivial configuration of a bulk scalar field. 
To illustrate its features, we work with a toy model 
in which the bulk is four dimensions. 
Using the superconformal formulation of SUGRA, 
we provide a systematic method of deriving the 3D effective action 
expressed by the superfields, 
which can basically be extended to 5D SUGRA straightforwardly.  
}}

\end{center}

\end{titlepage}

\clearpage

\section{Introduction}
Supersymmetry (SUSY) is one of the most promising candidates 
for the extension of the standard model. 
When you construct a realistic SUSY model, some mechanism 
for SUSY breaking is indispensable because no superparticles have been 
observed yet. 
However, in four dimensions, somewhat complicated set-up is necessary 
to break SUSY \cite{witten}. 

On the other hand, the brane-world scenario \cite{ADD,RS} 
has been attracted much attention and investigated in various frameworks. 
For example, some models can solve the gauge hierarchy problem \cite{RS}, 
and some can provide various patterns of gauge symmetry breaking 
without the Higgs fields \cite{kawamura}. 
Especially, the introduction of extra dimensions can provide 
a new SUSY breaking mechanism \cite{SS,STT}. 

When we construct the brane-world models, 
the radii of the extra dimensions have to be stabilized. 
One of the popular stabilization mechanisms is proposed in Ref.~\cite{GW}, 
and similar mechanisms have also been studied \cite{dewolfe}. 
These mechanisms involve a bulk scalar field 
that has a nontrivial field configuration.  
In such a case, the background geometry receives the backreaction 
of the scalar configuration. 
However, effects of such backreaction have been neglected 
in most works. 
In this paper, we will concentrate on such effects and 
investigate the SUSY breaking effects mediated through 
the deformation of the space-time geometry due to the backreaction. 

Since we would like to focus on the effects of SUSY breaking 
through the geometry, 
we will consider a situation where 
some scalar field has a {\it non-BPS} configuration in the hidden sector, 
which decouples from our visible sector. 
Then, the dominant contribution to SUSY breaking in the visible sector 
comes from the geometry-mediated effects. 
In this paper, we will suppose that 
there is only one extra dimension compactified on $S^1$ or $S^1/Z_2$, 
for simplicity

In order to understand qualitative features of this type of scenario, 
we will work with a simplified toy model, in which 
the bulk space-time is four dimensions (4D) 
and the effective theory is three-dimensional (3D). 
Although the 4D SUSY theories are less restrictive than the 5D ones 
and there are some different points between them, 
both situations have many common features. 
So we believe that our work is quite instructive to understand 
the characteristic features of the {\it geometry mediated SUSY breaking}. 
 
An interesting example of the stabilization mechanisms 
is proposed in Ref.~\cite{EMS}. 
In this article, the authors found a non-BPS solution 
in the 4D supergravity (SUGRA), which stabilizes the radius 
of the extra dimension and simultaneously generates 
a warped geometry. 

In the following, we will assume the existence of the non-BPS solution 
in 4D SUGRA, and derive the 3D effective theory on that background. 
Because the superfield formalism is very useful to discuss 
the phenomenology of the effective theory, 
we would like to express the 3D effective action in terms of 
3D superfields and the SUSY breaking terms. 
The main obstacle to our purpose is 
how we should define SUSY on the non-BPS background space-time. 
To solve this difficulty, the superconformal formulation of 
4D SUGRA~\cite{KU} is useful. 
Our strategy is as follows. 
We will see that 3D {\it global} SUSY can be defined 
on the gravitational background 
{\it before} the gauge fixing of the superconformal symmetry, 
even in the case that the background is non-BPS. 
So we can write an invariant action under the superconformal 
transformations in terms of 3D superfields constructed by 
that global SUSY. 
After the gauge fixing, the global SUSY will be broken 
and the SUSY breaking terms will emerge. 
In other words, {\it SUSY is broken by the gauge fixing conditions} 
in our method. 

The paper is organized as follows. 
In the next section, we will briefly review the superconformal approach 
of 4D SUGRA that is useful for our discussion. 
Then, we will rewrite the invariant action 
in terms of 3D superfields. 
In Sect.~\ref{SUSYbk}, we will impose the gauge fixing conditions 
and derive the 3D effective action with the SUSY breaking terms. 
Sect.~\ref{summary} is devoted to the summary and some discussions. 
Notations we use in this paper are listed in Appendix~\ref{notations}, 
and a brief comment on the radion superfield 
is provided in Appendix~\ref{radion_sf}.

\section{Action formula}
\subsection{Superconformal approach of 4D SUGRA}
We will consider a 4D $\cN=1$ supergravity model as a bulk theory. 
For our purpose, 
the superconformal approach \cite{KU} is useful. 
The field contents are as follows. 
\begin{description}
\item[Superconformal gauge fields:]
\be
  h_\mu = -\ge{\mu}{a}P_a+\frac{1}{2}\gomg{\mu}{ab}M_{ab}
 +\frac{\kp}{2}\brkt{\psi_\mu Q+\bar{\psi}_\mu\bar{Q}}+b_\mu D+A_\mu A 
 +\vph_\mu S+\bar{\vph}_\mu\bar{S}+\gf{\mu}{a}K_a, 
 \label{SCgauge_fd}
\ee
where $P_a,M_{ab},Q_\alp,\cdots$ are the generators of 
the 4D $\cN=1$ superconformal algebra\footnote{
The minus sign of the first term is necessary to match our notations 
to those of Ref.\cite{WB}. 
}, whose commutation relations are listed in Appendix~\ref{scSUGRA}. 
A constant~$\kp$ will be identified with the gravitational coupling 
after the gauge fixing~(\ref{scGF}), 
but it is just a constant at this stage. 
Throughout this paper, we will use $a,b,c,\cdots$ 
for the local Lorentz indices, 
and $\mu,\nu,\rho,\cdots$ for the world vector indices. 
\end{description}

Other fields form the superconformal multiplets, 
which are characterized by the Weyl weight~$w$ and 
the chiral weight~$n$.\footnote{
The definition of $n$ is different from that of Ref.~\cite{KU,KO} 
by a factor of $\frac{3}{2}$. 
} 
\begin{description}
\item[Chiral multiplets $(w=\frac{3}{2}n)$:]
\be
 \Phi^I=[\cA^I,\chi^I_\alp,\cF^I], \;\;\;\;\;(I=0,1,2,\cdots) 
\ee
where $\Sgm\equiv\Phi^{I=0}$ is the compensator multiplet 
whose Weyl weight is one, while $\Phi^{I\neq 0}$ have zero Weyl weights. 
We will assume that $\Phi^{I=1}$ is a field in the hidden sector, 
and its scalar component~$\phi^{\rm h}\equiv\cA^{I=1}$ 
has a nontrivial background configuration~$\phcl$.  
The two multiplets~$\Phi^{I=0,1}$ are neutral 
under the gauge group mentioned below. 
We will refer to the rest chiral multiplets~$\Phi^i\equiv\Phi^I$ 
($i=I=2,3,\cdots$) as the matter multiplets, 
and they may be charged under the gauge group. 

\item[(Internal) Gauge multiplet:] \mbox{}\\
The gauge multiplet~$[B^g_\mu,\lmd^g_\alp,D^g]$ can be embedded into 
a real general multiplet~$V^g$ with $w=n=0$ as 
\be
 V^g=[C^g,\zeta^g_\alp,\cH^g,\ge{a}{\mu}B^g_\mu
 -\frac{\kp}{2}(\psi_a\zeta^g+\bar{\psi}_a\bar{\zeta}^g),\lmd^g_\alp,D^g], 
\ee
where $\ge{a}{\mu}$ is an inverse matrix of the vierbein~$\ge{\mu}{a}$. 
Using the gauge degrees of freedom, 
we can set $C^g$, $\zeta^g_\alp$ and $\cH^g$ to zero\footnote{
This corresponds to the Wess-Zumino gauge in the global SUSY gauge theory.
}.
For simplicity, we will assume that the gauge group is abelian. 
In order for this gauge symmetry to remain in the 3D effective theory, 
it must be represented by a real representation 
because the 3D supermultiplets are real. 
Thus, we will concentrate on the case that 
the charged matter fields are doublets for the $SO(2)$ gauge group, 
and the gauge multiplet is represented as $2\times 2$ matrices on them. 
From the above gauge multiplet, we can construct the superfield strength 
multiplet~$\cW^g_\alp$, which is a chiral multiplet with 
an external spinor index~$\alp$, as 
\bea
 \cW^g_\alp=\sbk{-i\lmd^g_\alp,
  \frac{1}{\sqrt{2}}\brc{\dlt_\alp^{\;\bt}D^g-i\brkt{\sgm^{ab}}_\alp^{\;\;\bt}
  \hat{F}_{ab}^g}, \brkt{\sgm^a\cD_a\bar{\lmd}^g}_\alp}, 
\eea
where 
\bea
 \hat{F}_{\mu\nu}^g \defa \der_\mu B^g_\nu
  -i\brkt{\psi_\mu\sgm_\nu\bar{\lmd}^g+\bar{\psi}_\mu\bsgm_\nu\lmd^g}
  -(\mu\exch\nu), 
 \nonumber\\
 \cD_\mu\lmd^g_\alp \eql \der_\mu\lmd^g_\alp+\frac{1}{2}\gomg{\mu}{ab}
  \brkt{\sgm_{ab}\lmd^g}_\alp-i\psi_{\mu\alp}D^g-\brkt{\sgm^{ab}\psi_\mu}_\alp
  \hat{F}_{ab}^g-\frac{3}{2}b_\mu\lmd^g_\alp+iA_\mu\lmd^g_\alp. 
\eea
\end{description}
The multiplication laws of the above multiplets can be read off 
from Ref.~\cite{KO}. 
Some of them are listed in Appendix~\ref{scSUGRA}. 

In order to identify $P_a$ with the generator of 
the general coordinate transformation, some constraints are imposed 
on the field strengths of the gauge fields in Eq.(\ref{SCgauge_fd}). 
Then, $\gomg{\mu}{ab}$, $\vph_\mu$ and $\gf{\mu}{a}$ becomes 
dependent fields, \ie, 
they can be expressed in terms of $\ge{\mu}{a}$, $\psi_\mu^\alp$, $A_\mu$ 
and $b_\mu$. 
This corresponds to a deformation of the original superconformal algebra. 

We can construct an invariant action 
under the deformed superconformal transformation 
by using the action formulae, 
which are listed in Appendix~\ref{scSUGRA}. 
\be
 S = \int\dr^4x\sbk{G(\bar{\Phi},\Phi)+\Gm(\bar{\Phi},\Phi,V^g)}_D
  +\int\dr^4x\sbk{W(\Phi)}_F
  +\int\dr^4x\sbk{\frac{1}{4}\tr\brkt{\cW^{g\alp}\cW^g_\alp}}_F, 
 \label{4Daction_fml}
\ee
where $\Gm$ is a function determined 
so that $G+\Gm$ is $SO(2)$-gauge invariant\footnote{
Determination of $\Gm$ is explained in Ref.~\cite{WB}. 
}, and 
\bea
 G(\bar{\Phi},\Phi) \defa -\bar{\Sgm}\Sgm
  \exp\brc{-\frac{\kp^2}{3}\sum_{I\neq 0}
  \bar{\Phi}^{\bar{I}}\Phi^I}, 
 \nonumber\\
 W(\Phi) \defa \brkt{\frac{2}{3}}^{3/2}\kp^3\Sgm^3P(\Phi). 
 \label{def_fcn}
\eea
Here, we have assumed that the K\"{a}hler potential is minimal, 
and the superpotential~$P(\Phi)$ has a form: 
\be
 P(\Phi)=P_{\rm hid}(\Phi^{\rm h})+P_{\rm vis}(\Phi^i).
\ee
This corresponds to the assumption that the visible sector is decoupled 
from the hidden sector. 
In Eq.(\ref{4Daction_fml}), the physical fields~$\Phi^I$ ($I\neq 0$) 
and the gravitino~$\psi_\mu$ will be canonically normalized ones 
after the following gauge fixing.  

Finally, we can obtain the Poincar\'{e} SUGRA action 
by fixing the gauge of the superconformal symmetry, 
\bea
 \cA^\Sgm \eql \sqrt{\frac{3}{2}}\kp^{-1}
  \exp\brc{\frac{\kp^2}{6}\sum_{I\neq 0}\bar{\phi}^{\bar{I}}\phi^I}, 
  \;\;\;\brkt{\mbox{$D$, $A$-gauge fixing}}
 \nonumber\\ 
 \chi^\Sgm_\alp \eql \frac{\kp^2}{3}\cA^\Sgm\sum_{I\neq 0}
  \bar{\phi}^{\bar{I}}\chi^I_\alp, 
  \hspace{25mm}\brkt{\mbox{$S$-gauge fixing}}
 \nonumber\\
 b_\mu \eql 0. \hspace{49mm}\brkt{\mbox{$K$-gauge fixing}}
 \label{scGF}
\eea
By this gauge fixing, $\kp$ is identified with the gravitational coupling, 
\ie, the inverse of the 4D Planck mass~$\Mp$.  
The derivation of these gauge fixing conditions is briefly commented
in Appendix~\ref{scSUGRA}. 
After integrating out the auxiliary fields, 
we can check that the resulting action is identical 
to the one in Ref.~\cite{WB}. 

\subsection{Action on the gravitational background}
Throughout the paper, we will choose the $x_2$-direction 
as the extra dimension.  
So, $m,n=0,1,3$ denote the 3D vector indices and $y\equiv x_2$ is 
the coordinate of the extra dimension compactified on $S^1$ (or $S^1/Z_2$). 
We will use the letters with underbar for the local Lorentz indices, 
and those without the underbar for the world vector indices. 

The background metric is assumed to be of the form: 
\be
 ds^2=g_{\mu\nu}dx^\mu dx^\nu =e^{2A(y)}\eta_{\udl{m}\udl{n}}dx^m dx^n+dy^2. 
\ee

Since we are not interested in the gravitational interactions 
in the 3D effective theory, 
we will freeze the gravitational multiplet to its background value 
in the following discussion. 
\bea
 \vev{\ge{\mu}{a}} \eql \diag\brkt{e^A,e^A,1,e^A}, \nonumber\\
 \vev{\psi_\mu^\alp} \eql 0, \nonumber\\
 \vev{A_m} \eql 0, \nonumber\\
 \vev{A_y} \eql \frac{\kp^2}{2}\Im\brkt{\bphcl\dot{\phi}_{\rm bg}^{\rm h}}, 
 \label{grv_bg}
\eea
where the dot denotes the derivative with respective to $y$. 
In the last equation, we have used the equation of motion for $A_y$. 
Note that $A_\mu$, the gauge field of $U(1)_A$, is an auxiliary field 
since it does not have a kinetic term. 
Furthermore, we will drop the dependence on 
$b_\mu$, the gauge field of the dilatation~$D$, in the following calculation
because it will eventually be set to zero 
by the gauge fixing condition in Eq.(\ref{scGF}). 


Then, the action~(\ref{4Daction_fml}) becomes 
\be
 S_{\rm bg}=\int\dr^4x\; e\sbk{\frac{C}{3}R+D
  +\brc{\cF+\frac{1}{4}\tr\cF_\cW+\hc}}. 
 \label{S_on_bg}
\ee
Here, $e\equiv\det(\ge{\mu}{a})$ and the Ricci scalar~$R$ are their background 
values. 
\bea
 e \eql e^{3A}, \nonumber\\
 R \eql 6(\ddot{A}+2\dot{A}^2).  \label{bg_Ricci}
\eea
From the multiplication laws of the superconformal multiplets, 
we can compute each components as 
\bea
 C \eql G(\bar{\cA},\cA), \nonumber\\
 D \eql G_{I\bar{J}}\brc{-2\cD^{\mu}\bar{\cA}^{\bar{J}}\cD_\mu\cA^I
  -i\chi^I\sgm^a\cD_a\bar{\chi}^{\bar{J}}
  -i\bar{\chi}^{\bar{J}}\bsgm^a\cD_a\chi^I+2\bar{\cF}^{\bar{J}}\cF^I}
 \nonumber\\
  &&+G_{IJ\bar{K}}\brc{i\chi^J\sgm^a\bar{\chi}^{\bar{K}}\cD_a\cA^I
  -\bar{\cF}^{\bar{K}}\brkt{\chi^I\chi^J}} 
 \nonumber\\
  &&+G_{I\bar{J}\bar{K}}\brc{i\bar{\chi}^{\bar{K}}\bsgm^a\chi^I
  \cD_a\bar{\cA}^{\bar{J}}
  -\cF^I\brkt{\bar{\chi}^{\bar{J}}\bar{\chi}^{\bar{K}}}}
  +\frac{1}{2}G_{IJ\bar{K}\bar{L}}\brkt{\chi^I\chi^J}
  \brkt{\bar{\chi}^{\bar{K}}\bar{\chi}^{\bar{L}}}
 \nonumber\\
  &&+2\sqrt{2}i\brc{\Dkl_I\lmd^g\chi^I
  -\Dkl_{\bar{I}}\bar{\lmd}^g\bar{\chi}^{\bar{J}}}+2\Dkl D^g, 
 \nonumber\\
 \cF \eql W_I\cF^I-\frac{1}{2}W_{IJ}\chi^I\chi^J, 
 \nonumber\\
 \cF_\cW \eql \brkt{D^g}^2-\frac{1}{2}F^{g\mu\nu}F_{\mu\nu}^g
  -\frac{i}{4}\ep^{\mu\nu\rho\sgm}F_{\mu\nu}^gF_{\rho\sgm}^g
  -2i\lmd^g\sgm^a\cD_a\bar{\lmd}^g, 
 \label{components}
\eea
where $G(\bar{\cA},\cA)$ and $W(\cA)$ are functions 
defined in Eq.(\ref{def_fcn}), 
$F_{\mu\nu}^g\equiv \der_\mu B^g_\nu-\der_\nu B^g_\mu$ is 
the ordinary field strength, 
and $\Dkl$ is the Killing potential for the $SO(2)$ isometry 
of the K\"{a}hler manifold determined by $G(\bar{\cA},\cA)$ \cite{WB}. 
The subscripts~$I,J,\cdots$ denote the derivatives with respective to 
the corresponding scalar fields. 
The covariant derivatives are defined as  
\bea
 \cD_\mu\cA^I \defa \der_\mu\cA^I+igB^g_\mu\cA^I
 \nonumber\\
 \cD_\mu\chi^I_\alp \defa \der_\mu\chi^I_\alp+igB^g_\mu\chi^I_\alp
  +\frac{1}{2}\gomg{\mu}{ab}\brkt{\sgm_{ab}\chi^I}_\alp, 
\eea
where $g$ represents the corresponding charge for the $SO(2)$ gauge group, 
and $\gomg{\mu}{ab}$ is the spin connection. 
Here, we have taken the Wess-Zumino gauge.

\subsection{Global SUSY transformation and superfields}
We will rewrite the action~(\ref{S_on_bg}) on the 3D $\cN=1$ 
superspace~$(x^m,y,\tht^\alp)$, 
where $\tht$ is a Grassmannian coordinate 
that is a 3D Majorana spinor. 

\subsubsection{The case of $\vev{A_y}=0$} \label{case_Ay0}
First, we will consider the case of $\vev{A_y}=0$. 
Let us consider a superconformal transformation~$\dQ$ 
that  preserves the gravitational background~(\ref{grv_bg}). 
Namely, 
\be
 \dQ=\dlt_Q(\ep)+\dlt_D(\lmd_D)+\dlt_A(\vtht_A)+\dlt_S(\eta)+\dlt_K(\xi_K). 
 \label{sc_dlt}
\ee
with 
\bea
 \ep_\alp \eql \frac{ie^{i\vtht_0}}{\sqrt{2}}e^{\frac{A}{2}}\ep_{0\alp}, 
 \nonumber\\
 \eta_\alp \eql \frac{1}{2}\dot{A}\brkt{\sgm^2\bar{\ep}}_\alp, \nonumber\\
 \lmd_D \eql \vtht_A=\xi_{K\mu}=0, 
 \label{gl_prmt}
\eea
where $\ep_0$ is a constant 3D Majorana spinor ($(\ep_0^\alp)^*=\ep_0^\alp$), 
and $\vtht_0$ is a constant phase.  
The transformation law of the gravitational multiplet is listed 
in Appendix~\ref{scSUGRA}. 

We can easily see that $\dQ(\ep_0)$ satisfies the 3D $\cN=1$ SUSY algebra. 
So we can identify $\dQ(\ep_0)$ with {\it global} 3D SUSY transformation. 
Since $\dQ(\ep_0)$ preserves the gravitational background, 
the action~$S_{\rm bg}$ in Eq.(\ref{S_on_bg}) is invariant under it. 
This ensures that $S_{\rm bg}$ can be rewritten in terms of 
the superfields constructed by $\dQ(\ep_0)$. 

%
%

We can construct superfields as follows. 
\bea
 \vph^I \defa e^{\dQ(\tht)}\cA^I, \nonumber\\
 \rho_\alp \defa e^{\dQ(\tht)}\zeta^g_{1\alp}, \nonumber\\
 \sgm \defa e^{\dQ(\tht)}\frac{1}{2}(-B^g_y+M^g), 
\eea
where 3D Majorana spinor~$\zeta^g_1$ and the real scalar~$M^g$ are defined as 
\bea
 \zeta^{g\alp} \defa \frac{e^{-i\vtht_0}}{\sqrt{2}}
  \brkt{\zeta^{g\alp}_1+i\zeta^{g\alp}_2}, \nonumber\\
 \cH^g \defa \frac{e^{-2i\vtht_0}}{2}\brkt{M^g+iN^g}.  
 \label{decomp_zH}
\eea
It is easy to check that a global transformation induced by 
the differential operator:
\be
 \hat{Q}_\alp=\frac{\der}{\der\tht^\alp}+i\brkt{\gm_{(3)}^{\udl{m}}\tht}_\alp 
  \der_m 
\ee
is identical to the above defined global SUSY transformation~$\dQ(\ep_0)$. 

We can also construct a superfield~$e^{\dQ(\tht)}C^g$, but this is completely 
eliminated after taking the Wess-Zumino gauge and 
the further 4D gauge transformation \cite{sakamura1}.  
In this gauge, we can obtain the following expressions. 
\bea
 \vph^I(x,\tht) \defa \cA^I+e^{i\vtht_0}e^{\frac{A}{2}}\tht\chi^I
  +\frac{e^A}{2}\tht^2 f^I, 
 \nonumber\\
 \rho_\alp(x,\tht) \defa -ie^{\frac{A}{2}}\brkt{\gm_{(3)}^m\tht}_\alp B^g_m
  -e^A\tht^2\lmd^g_{2\alp}, \nonumber\\
 \sgm(x,\tht) \defa -B^g_y+e^{\frac{A}{2}}\tht\lmd^g_1
  -\frac{1}{2}e^A\tht^2 D^g, 
 \label{def_sf}
\eea
where $\gm_{(3)}^m\equiv \ge{\udl{n}}{m}\gm_{(3)}^{\udl{n}}$, 
\be
 f^I \equiv i\brkt{\der_y\cA^I+igB^g_y\cA^I+e^{2i\vtht_0}\cF^I+w\dot{A}\cA^I},
  \;\;\;\;\;(\mbox{$w$: Weyl weight})
 \label{def_f}
\ee
and 3D Majorana spinors~$\lmd_1^g$ and $\lmd_2^g$ are defined as 
\be
 \lmd^{g\alp} \equiv \frac{e^{i\vtht_0}}{\sqrt{2}}
  \brkt{\lmd_1^{g\alp}+i\lmd_2^{g\alp}}.  
 \label{decomp_lmd}
\ee
Here and hereafter, we will use the 3D notations for spinors, 
which are listed in Appendix~\ref{3Dnotations}. 
The superfield~$\vph^I$, which is constructed from the chiral multiplet, 
is a 3D scalar superfield, and 
$\rho$ and $\sgm$ are the 3D gauge superfield and 
the gauge scalar superfield, respectively. 
After fixing the background as Eq.(\ref{grv_bg}), 
the action still has the 3D supergauge symmetry. 
Under its transformation~$\dlt^{\rm sg}$, 
the above superfields are transformed as  
\bea
 \dlt^{\rm sg}\vph^I \eql ig\Omega\vph^I, \nonumber\\
 \dlt^{\rm sg}\rho_\alp \eql -e^{-\frac{A}{2}}D_{\tht\alp}\Omega, \nonumber\\
 \dlt^{\rm sg}\sgm \eql \der_y\Omega, 
 \label{3Dgauge_trf}
\eea
where 
\be
 D_{\tht\alp}\equiv\frac{\der}{\der\tht^\alp}
  -i\brkt{\gm_{(3)}^{\udl{m}}\tht}_\alp\der_m, 
\ee
and the transformation parameter~$\Omega$ is a 3D real scalar superfield. 
From these transformation properties, 
we can construct the following supergauge covariant derivatives. 
\bea
 D_\alp \defa D_{\tht\alp}+ige^{\frac{A}{2}}\rho_\alp,
  \nonumber\\
 \nabla_y \defa \der_y-ig\sgm. 
\eea
The gauge invariant quantities under Eq.(\ref{3Dgauge_trf}) are 
\bea
 w^g_\alp \defa \frac{1}{4}e^{-A}D_\tht^2\rho_\alp
  +\frac{i}{2}\brkt{\gm_{(3)}^m\der_m\rho}_\alp \nonumber\\
  \eql \lmd^g_{2\alp}+e^{\frac{A}{2}}\brkt{\gm_{(3)}^{mn}\tht}_\alp
  F_{mn}^g-\frac{i}{2}e^A\tht^2\brkt{\gm_{(3)}^m\der_m\lmd^g_2}_\alp, 
  \nonumber\\
 u^g_\alp \defa e^{-\frac{A}{2}}D_{\tht\alp}\sgm
  +\brkt{\der_y+\frac{\dot{A}}{2}}\rho_\alp \nonumber\\
  \eql \lmd^g_{1\alp}-e^{\frac{A}{2}}\tht_\alp D^g
  +ie^{\frac{A}{2}}\brkt{\gm_{(3)}^m\tht}_\alp
  \brkt{\der_m B^g_y-\der_y B^g_m} 
  \nonumber\\
  &&+e^A\tht^2\brc{\frac{i}{2}\brkt{\gm_{(3)}^m\der_m\lmd^g_1}_\alp
  -\brkt{\der_y+\frac{3}{2}\dot{A}}\lmd^g_{2\alp}}. 
 \label{def_wu}
\eea
Here, $w^g$ is the 3D superfield strength. 

Using these superfields, we can rewrite the action~(\ref{S_on_bg}) as 
\bea
 S_{\rm bg} \eql \int\dr^4x\int\dr^2\tht \left[
  2e^AG_{I\bar{J}}D^\alp\bar{\vph}^{\bar{J}}D_\alp\vph^I
  +4e^{2A}\Im\brc{G_I\nabla_y\vph^I+e^{-2i\vtht_0}W(\vph)} \right.
 \nonumber\\
  &&\hspace{22mm}\left. 
  +\frac{e^{2A}}{2}\tr\brc{\brkt{u^g}^2-\brkt{w^g}^2}\right].  
 \label{action_fml1}
\eea
This corresponds to an extended version of the formula 
in Ref.~\cite{sakamura1} to the warped geometry. 
We can explicitly check that the above expression reproduces 
Eq.(\ref{S_on_bg}) with Eqs.(\ref{bg_Ricci}), (\ref{components}) 
up to total derivative terms by performing the $\tht$-integration. 

It should be noted that the above action is the expression 
{\it before} the superconformal gauge fixing~(\ref{scGF}). 
Thus, it is not yet the SUGRA action at this stage. 


\subsubsection{The case of $\vev{A_y}\neq 0$}
Now, we will extend Eq.(\ref{action_fml1}) to the case that 
$\vev{A_y}\neq 0$. 
Notice that we can move to the gauge where $\vev{A_y}=0$ 
by using the $U(1)_A$-transformation, 
while keeping the other backgrounds unchanged. 
In this gauge, each field is rotated by its chiral weight 
with the parameter:
\be
 \vtht_A(y)=-\int^y \dr z\;\vev{A_y}(z). \label{def_xiA}
\ee 
An integration constant can be absorbed into the constant phase~$\vtht_0$. 
If we reconstruct 3D superfields by using the rotated bulk fields, 
the expression of $S_{\rm bg}$ becomes the same form 
as Eq.(\ref{action_fml1}). 
Thus, we can obtain the desired $\vev{A_y}$-dependent action 
by re-expressing each component field in Eq.(\ref{action_fml1}) in terms of 
the original ones, namely, 
by replacing each fields in Eq.(\ref{action_fml1}) 
with the rotated ones with the parameter~$\vtht_A$. 

The rotated scalar superfield is  
\be
 \vph^I = e^{-\frac{2w}{3}i\vtht_A}\brc{
  \cA^I+e^{i\vtht_A}e^{\frac{A}{2}}\tht\chi^I
  +\frac{i}{2}e^A\tht^2\brkt{\der_y\cA^I+igB^g_y\cA^I
  +\frac{2w}{3}i\vev{A_y}\cA^I 
  +e^{2i\vtht_A}\cF^I+w\dot{A}\cA^I}}. 
 \label{rc_scalar}
\ee
Here, note that the chiral weight of the chiral multiplet is equal 
to $\frac{2w}{3}$. 

Unlike the chiral multiplet\footnote{
The chiral multiplet is also decomposed into two real multiplets, 
but these are the same type of the multiplet, \ie, the scalar multiplet. 
So we can treat them as one complex scalar multiplet. 
}, the 4D gauge multiplet must be 
decomposed into two different types of the 3D real multiplets 
to form the 3D superfields. 
So, the $U(1)_A$-rotation of the 4D complex field corresponds to 
the mixing between two 3D real fields. 
For example, $\lmd^g_1$ and $\lmd^g_2$ are mixed 
under the $U(1)_A$-rotation. 
This mixing breaks the global SUSY because they belong to 
different supermultiplets. 
In fact, since the mixing angle~$\vtht_A$ is $y$-dependent, 
the gaugino mass term appears from the $y$-derivative 
in the definition of $u^g$ in Eq.(\ref{def_wu}). 
This is very similar to the Scherk-Schwarz (SS) SUSY breaking 
in the flat space-time \cite{SS}. 
However, SUSY breaking due to the non-zero $\vev{A_y}$ are different 
from the SS breaking in the following points. 
First, $U(1)_A$ is not a symmetry of the theory 
after the gauge fixing~(\ref{scGF}). 
So it is independent of $U(1)_R$ symmetry that is relevant to the SS breaking. 
Second, $\vev{A_y}$ is not an input parameter as in the SS mechanism, 
but is determined by the hidden sector dynamics, 
which also induces 
another types of SUSY breaking terms through the deformation of the warp 
factor as will be seen in Sect.~\ref{kp_ex}. 


As a result, Eq.(\ref{action_fml1}) is modified as 
\bea
 S_{\rm bg} \eql \int\dr^4 x\int\dr^2\tht\left[
  2e^AG_{I\bar{J}}D^\alp\bar{\vph}^{\bar{J}}D_\alp\vph^I
  +4e^{2A}\Im\brc{G_I\nabla_y\vph^I
  +e^{-2i\tht_A}W(\vph)} \right. \nonumber\\
 &&\hspace{21mm}\left.
  +\frac{8}{3}e^{2A}\vev{A_y}G
  +\frac{e^{2A}}{2}\tr\brc{T^{-1}\brkt{u^g}^2-T\brkt{w^g}^2}\right], 
 \label{action_fml2}
\eea
where $\vph^I$ is redefined as 
\be
 \vph^I\equiv \cA^I+e^{i\vtht_A}e^{\frac{A}{2}}\tht\chi^I
  +\frac{i}{2}e^A\tht^2\brkt{\der_y\cA^I+igB^g_y\cA^I
  +\frac{2w}{3}i\vev{A_y}\cA^I 
  +e^{2i\vtht_A}\cF^I+w\dot{A}\cA^I}, 
\ee
and 
\be
 T\equiv 1+e^A\tht^2\vev{A_y} 
\ee 
is a spurion superfield, which can be interpreted as a background value 
of the radion superfield mentioned in Appendix~\ref{radion_sf}. 

Eq.(\ref{action_fml2}) has a similar form to the expression in Ref.~\cite{MP}, 
but note that the latter is a result {\it after} 
the superconformal gauge fixing.

\section{Gauge fixing and SUSY breaking}\label{SUSYbk}
\subsection{Deviation from SUSY limit}
To discuss SUSY breaking, we should consider the supersymmetric limit 
first. 
When the background is a BPS configuration, a half of the bulk SUSY 
is conserved. 
In this case, the direction of the conserved SUSY is parametrized 
by the Killing spinor~$\ep^\alp(y)$, 
which is determined by the background configuration. 
Here, we will parametrize the Killing spinor as 
\be
 \ep^\alp=ie^{i\vtht}\abs{\ep^\alp}. 
\ee 

The BPS equations are obtained as follows by requiring 
the local SUSY transformation\footnote{
The local SUSY transformation is obtained by requiring that 
it preserves the gauge fixing conditions~(\ref{scGF}). 
} of the fermionic fields to vanish \cite{EMS}. 
\bea
 \dot{A} \eql 
  -\kp^2 e^{-2i\vtht}e^{\frac{\kp^2}{2}\abs{\phi^{\rm h}}^2}P_{\rm hid}, 
 \nonumber\\
 \dot{\phi}^{\rm h} \eql e^{2i\vtht}e^{\frac{\kp^2}{2}\abs{\phi^{\rm h}}^2}
  \brkt{\frac{\der\bar{P}_{\rm hid}}{\der\bar{\phi}^{\rm h}}
  +\kp^2\phcl \bar{P}_{\rm hid}} \nonumber\\
 \dot{\abs{\ep^\alp}} \eql \frac{\dot{A}}{2}\abs{\ep^\alp}, \nonumber\\
 \dot{\vtht} \eql -\frac{\kp^2}{2}\Im\brkt{\bar{\phi}^{\rm h}
  \dot{\phi}^{\rm h}}. 
 \label{BPSeq}
\eea
Here, we have supposed that only $\phi^{\rm h}$ has a nontrivial field 
configuration among the physical scalar fields. 

By solving the third equation, the Killing spinor can be written as 
\be
 \ep_\alp=\frac{ie^{i\vtht}}{\sqrt{2}}e^{\frac{A}{2}}\ep_{0\alp},  
\ee
where $\ep_0$ is a constant 3D Majorana spinor. 
Note that R.H.S. of the last equation in Eq.(\ref{BPSeq}) is equal 
to $-\vev{A_y}$. (See Eq.(\ref{grv_bg}).)
This means that $\vtht_A$ in Eq.(\ref{def_xiA}) can be interpreted as 
the phase of the Killing spinor~$\vtht$ when the background is BPS. 
Thus, from the first two equations in Eq.(\ref{BPSeq}), 
we can define the following two functions 
to characterize the deviation from the SUSY limit. 
\bea
 \fG \eql \frac{\dot{A}}{\kp^2}+e^{-2i\vtht_A}e^{\frac{\kp^2}{2}\abs{\phcl}^2}
  \bar{P}_{\rm hid},  \nonumber\\
 \fhcl \eql \dot{\phi}_{\rm bg}^{\rm h}
  -e^{2i\vtht_A}e^{\frac{\kp^2}{2}\abs{\phcl}^2}
  \brc{\frac{\der\bar{P}_{\rm hid}}{\der\bar{\phi}^{\rm h}}(\phcl)
  +\kp^2\phcl\bar{P}_{\rm hid}(\phcl)}. 
 \label{od_pm}
\eea
From Eq.(\ref{action_fml2}), we can see that $\vev{A_y}$ also 
characterizes the magnitude of SUSY breaking. 
Therefore, we will refer to these three functions 
as the {\it SUSY-breaking functions} in the following.

\subsection{Gauge fixing and $\kp$-expansion} \label{kp_ex}
To obtain the action, we have to fix the redundant superconformal 
gauge symmetry. 
However, the gauge fixing conditions~(\ref{scGF}) are not invariant 
under the global SUSY transformation~$\dQ(\ep_0)$. 
Thus it will be broken after the gauge fixing. 

Practically, it is enough to calculate the action 
up to the next leading order in the $\kp$-expansion. 
In order to count the power of $\kp$ in the calculation, 
it is convenient to decompose $\vph^\Sgm$ as 
\be
 \vph^\Sgm=\Abg+\tl{\vph}^\Sgm, 
\ee
where 
\be
 \Abg\equiv\sqrt{\frac{3}{2}}\kp^{-1}e^{\frac{\kp^2}{6}\abs{\phcl}^2} 
\ee
is the background value of $\cA^\Sgm$. 
Then, only $\Abg$ reduces the power of $\kp$.  
\bea
 \Abg \eql \cO(\kp^{-1}), \nonumber\\
 \vph^{I\neq 0} \eql \cO(\kp^0), \nonumber\\
 \dot{\phi}_{\rm bg}^\Sgm,\tl{\vph}^\Sgm \eql \cO(\kp).  
\eea

The action is calculated from Eq.(\ref{action_fml2}) with 
\bea
 G \eql -\bar{\vph}^{\bar{\Sgm}}\vph^\Sgm
  \exp\brc{-\frac{\kp^2}{3}\sum_{I\neq 0}\bar{\vph}^{\bar{I}}\vph^I}, 
 \nonumber\\
 W \eql \brkt{\frac{2}{3}}^{3/2}\kp^3\brkt{\vph^\Sgm}^3
  \brc{P_{\rm hid}(\vph^{\rm h})+P_{\rm vis}(\vph^i)}, 
\eea 
and rewriting the components of $\tl{\vph}^\Sgm$ in terms of 
the physical fields by using the gauge fixing conditions~(\ref{scGF}). 

Since we are interested only in the visible sector, 
we will replace $\vph^{\rm h}$ with its background value 
\be
 \vph^{\rm h}_{\rm bg}=\phcl+\frac{e^A}{2}\tht^2 \fhcl, 
\ee
where $\fhcl$ is defined in Eq.(\ref{od_pm}), and ignore 
the fluctuation modes around it.  


Then, the resulting action in the visible sector is obtained as follows. 
\bea
 S_{\rm vis} \eql \int\dr^4x\int\dr^2\tht
 \left[e^A\sum_i D^\alp\bar{\vph}^{\bar{i}}D_\alp\vph^i
 +4e^{2A}\Im\brc{\frac{1}{2}\sum_i\bar{\vph}^{\bar{i}}\nabla_y\vph^i
 +e^{\frac{\kp^2}{2}\abs{\phcl}^2}P_{\rm vis}(\vph^i)}\right. \nonumber\\
 &&\hspace{20mm}\left.
 +\frac{4}{3}e^{2A}\vev{A_y}\sum_i\bar{\vph}^{\bar{i}}\vph^i\right] 
 \nonumber\\
 &&+\kp^2\int\dr^4x\; e^{3A}\left[\Im\brc{-2\sum_i\fG\phi^i\bar{f}^{\bar{i}}
 +6\hfbg P_{\rm vis}-\frac{4\vev{A_y}}{3\kp^2}\brkt{
 3P_{\rm vis}-\sum_i\phi^i\frac{\der P_{\rm vis}}{\der\phi^i}}}\right.
 \nonumber\\
 &&\hspace{26mm}\left.-\frac{\vev{A_y}}{2\kp^2}\tr\brc{
 \brkt{\lmd^g_1}^2+\brkt{\lmd^g_2}^2}+\cdots\right]
 +\cO(\kp^4), 
 \label{Smatter}
\eea
where the ellipsis in the last line denotes quartic or higher terms. 
The function~$\hfbg(y)$ is defined as  
\be
 \hfbg \equiv \frac{i}{3}\brc{\bphcl \fhcl+3\fG} 
\ee
and corresponds to the $f$-term of the background value of $\vph^\Sgm$, 
\be
 \vph_{\rm bg}^\Sgm = \Abg\brkt{1+\kp^2\frac{e^A}{2}\tht^2\hfbg}. 
\ee
Note that $\vev{A_y}=\cO(\kp^2)$. 
The action at the lowest order~$\cO(\kp^0)$ is manifestly supersymmetric 
because it is written only by the superfields. 
This is a trivial result since SUSY breaking effects are mediated 
only through the space-time geometry, \ie, the gravitational effect. 

Note that the gauge fixing conditions~(\ref{scGF}) 
is not invariant under $\dQ(\ep_0)$ 
even in the case that the background configuration is BPS. 
This reflects the fact that the SUGRA action cannot be expressed 
only by the superfields. 
So there are also the SUGRA terms besides the SUSY breaking terms 
among terms that cannot be written by the superfields in Eq.(\ref{Smatter}). 
Such SUGRA terms appear in the quartic or higher terms. 
Therefore, the quadratic and cubic terms in the third and fourth lines of 
Eq.(\ref{Smatter}) are purely SUSY breaking terms.  
These terms will be important as the soft SUSY-breaking terms 
when we extend our discussion to the 4D effective theory of 5D SUGRA.  

Note also that 
all SUSY breaking terms should be associated with at least one of 
the SUSY-breaking functions~$\fhcl$, $\fG$, $\vev{A_y}$.   
Therefore, we may pick up only terms involving such functions 
in order to obtain the SUSY-breaking terms. 
Other terms are ensured to be cancelled with each other. 
This simplifies somewhat tedious calculations 
in the gauge fixing procedure.

\subsection{Example of non-BPS configuration}
For a specific example of the non-BPS backgrounds, 
we will take a configuration discussed in Ref.~\cite{EMS}. 

\subsubsection{Properties of the background}
In Ref.~\cite{EMS}, a topological winding number is introduced 
to stabilize the non-BPS configuration 
along the $S^1$-compactified extra dimension.  
However, a periodic solution with no winding is also possible 
in the case that the extra dimension is compactified on $S^1/Z_2$. 
The configuration corresponds to the coexistence of a BPS domain wall at $y=0$ 
and an anti-BPS wall at $y=\pi r$. 
We will refer to the 3D hyperplanes at $y=0,\pi r$ 
as the ``boundaries'' despite the $S^1$-compactification 
as the authors of Ref.~\cite{EMS} do. 

The radius of the extra dimension~$r$ is stabilized 
by the scalar configuration, 
and its size is determined by the tensions on the 3D boundaries\footnote{
The model in Ref.~\cite{EMS} does not introduce any boundary fields, 
but admits positive and negative tensions on the boundaries. 
}. 
For example, for a unit winding solution, 
\be
 \pi r \sim \frac{1}{\Lmd}\ln\frac{\Lmd^4}{\kp^2\tau_0^2},  
\ee
where $\Lmd$ is a characteristic mass scale of the potential 
for $\phi^{\rm h}$, and $\tau_0$ is the boundary tension at $y=0$\footnote{
The tension on the other boundary~$\tau_{\pi r}$ must be negative and 
its magnitude is determined by $\tau_0$. 
The fine tuning between $\tau_0$ and $\tau_{\pi r}$ corresponds to 
the assumption of the vanishing cosmological constant 
in the 3D effective theory. 
}. 
Here, we have used the weak-gravity approximation \cite{EMS}. 
In this approximation, the warp factor~$A(y)$ is calculated as 
\be
 A(y)=A_0-\frac{\kp^2\tau_0}{4}y+\cdots, 
\ee
where $A_0$ is a normalization constant and the ellipsis denotes 
the higher-order corrections in $\kp$. 

In the limit of $r\to\infty$ (or $\tau_0\to 0$), 
the configuration around $y=0$ approaches to a BPS solution with $\vtht=0$ 
and 3D $\cN=1$ SUSY is recovered. 

In this model, the scalar configuration~$\phcl$ is real 
and thus $\vev{A_y}=0$. 
So the gaugino mass is not induced on this background.

\subsubsection{Mode expansion}
Since we are interested in the low-energy effective theory, 
we will focus on the zero-modes of the bulk fields. 
We will introduce the kink mass terms.  
\be
 P_{\rm vis}(\vph^i)=\sum_i m_i\varepsilon(y)(\vph^i)^2
 +(\mbox{interaction terms}),
\ee
where $m_i>0$ are the mass parameters and the function~$\varepsilon(y)$ is 
defined as 
\be
 \varepsilon(y)=\left\{\begin{array}{cl}1 & \mbox{for $0 < y <\pi r$}, \\
  0 & \mbox{for $y=0,\pm\pi r$}, \\  -1 & \mbox{for $-\pi r < y <0$}. 
  \end{array}\right. 
\ee
Here,we have assumed that $\vph^i$ is gauge-singlet for simplicity. 
The extension to the charged matter case is straightforward. 

Then, the fermionic component of $\vph^i$ is expanded as 
\be
 \chi^i(x^m,y)=\frac{1}{\sqrt{2}}\brc{\uR{0}(y)\chi_{{\rm R}0}^i(x^m)
  +i\uI{0}(y)\chi_{{\rm I}0}^i(x^m)}+(\mbox{massive modes}). 
 \label{chi_md_ex}
\ee
The mode-functions are obtained by solving the linearized 
equations of motion. 
For example, those of the zero-modes are 
\bea
 \uR{0}(y) \eql C_{\rm R}^ie^{-\frac{3}{2}A(y)}e^{-m_i\abs{y}}, \nonumber\\
 \uI{0}(y) \eql C_{\rm I}^ie^{-\frac{3}{2}A(y)}e^{m_i\abs{y}}, 
\eea
where $C_{\rm R}$ and $C_{\rm I}$ are the normalization constants 
determined by
\be
  \int\dr y\;e^{2A}\brkt{\uR{0}}^2 = \int\dr y\;e^{2A}\brkt{\uI{0}}^2=1. 
\ee
Depending on the value of $m_i$, the zero-mode~$\chi_{{\rm R}0}^i$ 
can be localized around $y=0$. 

As mentioned above, 3D $\cN=1$ SUSY characterized by $\vtht=0$ 
is a good symmetry around $y=0$ when the radius~$r$ is large. 
Thus, if $\uR{0}$ is localized around $y=0$, $\chi^i_{{\rm R}0}$ forms 
a supermultiplet for that SUSY. 
Namely, it is useful to define a 3D superfield:
\be
 \vph^i_0(x^m,\tht) = a^i_0(x^m)+\tht\chi^i_{{\rm R}0}(x^m)
  +\frac{1}{2}\tht^2 f^i_0(x^m). 
\ee

On the other hand, $\uI{0}$ is localized around $y=\pi r$, 
and an opposite half of the bulk SUSY 
characterized by $\vtht=-\pi$ is a good symmetry there.  
So $\chi^i_{{\rm I}0}$ forms a supermultiplet for this SUSY. 
\be
 \vph^{\prime i}_0(x^m,\tht^\prime) = 
  a^{\prime i}_0(x^m)+\tht^\prime\chi^i_{{\rm I}0}(x^m)
  +\frac{1}{2}\tht^{\prime 2}f^{\prime i}_0(x^m). 
\ee


Using the method used in Ref.~\cite{sakamura2}, 
we can see that the zero-modes are embedded into the bulk field~$\vph^i$ as 
\bea
 \vph^i(x^m,y,\tht) \eql \frac{e^{\frac{A}{2}}}{\sqrt{2}}
  \brc{\uR{0}(y)\vph^i_0(x^m,\tht)
  +\uI{0}(y+ie^A\tht_2^2)\vph^{\prime i}_0(x^m,i\tht)} \nonumber\\
  &&+(\mbox{massive modes}).  
 \label{zm_embed}
\eea

From this expression, 
\be
 \Im\brkt{\phi^i\bar{f}^i}=-\frac{e^{\frac{A}{2}}}{2}
  \brkt{\uR{0}\der_y\uI{0}}\phi^i_0\phi^{\prime i}_0 
  -\frac{e^{\frac{A}{2}}}{2}\uI{0}\der_y\uI{0}
  (\phi^{\prime i}_0)^2
  +\cdots, 
\ee
where the ellipsis denotes terms involving massive Kaluza-Klein modes. 
Thus, the first term in the third line of Eq.(\ref{Smatter}) 
gives mixing terms between the scalar modes localized on 
the different boundaries\footnote{
Note that $\fG$ is real in this model. 
}. 

Since we do not suppose any mechanism for localization of the gauge field, 
the mode expansion of the gauge supermultiplet is trivial. 
\bea
 \rho^\alp(x^m,y,\tht) \eql \frac{1}{\sqrt{2\pi r}}\rho^\alp_0(x^m,\tht)
 \nonumber\\
  &&+\sum_{n=1}^\infty\frac{1}{\sqrt{\pi r}}\brc{
  \cos\frac{ny}{r}\cdot\rho^\alp_{(n+)}(x^m,\tht)
  +\sin\frac{ny}{r}\cdot\rho^\alp_{(n-)}(x^m,\tht)}, \nonumber\\
 \sgm(x^m,y,\tht) \eql \frac{e^{-\frac{A}{2}}}{\sqrt{2\pi r}}\sgm_0(x^m,\tht)
 \nonumber\\
  &&+\sum_{n=1}^\infty\frac{e^{-\frac{A}{2}}}{\sqrt{\pi r}}\brc{
  \cos\frac{ny}{r}\cdot\sgm_{(n+)}(x^m,\tht)
  +\sin\frac{ny}{r}\cdot\sgm_{(n-)}(x^m,\tht)}. 
\eea
The signs in the label of the 3D superfields denote the parity 
under $y\to -y$. 
The normalization factors should be modified in the case of 
the orbifold compactification. 

Plugging these expressions into Eq.(\ref{Smatter}) and 
performing the $y$-integration, 
we can obtain the desired action for the visible sector.

\subsubsection{SUSY-breaking scales in the effective theory} \label{scales}
In this example, scales introduced in the bulk theory are 
the Planck scale~$\Mp$, the kink mass parameters~$m_i$, 
the characteristic scale of the hidden sector dynamics~$\Lmd$ 
and the compactification scale~$r^{-1}$ (or the tension scale~$\tau_0^{1/3}$). 
From Eq.(\ref{Smatter}), the SUSY-breaking scalar masses 
in the 3D effective theory have a form of 
\be
 m_S^2=\kp^2\Lmd^3m_i \alp(m_i,\Lmd,r), 
\ee
where $\alp(m_i,\Lmd,r)$ is a dimensionless function expressed 
by the overlap integral of the mode-functions and 
the SUSY-breaking functions~$\fhcl$, $\fG$. 
For large~$r$, 
its asymptotic form is 
\be
 \alp(m_i,\Lmd,r)\sim e^{-c(m_i)\Lmd r}, \label{asymp_alp}
\ee
where $c(m_i)$ is a positive number 
depending on the strength of the localization of the mode-functions. 

The gaugino mass~$m_g$ does not appear at tree-level. 
However, if the background configuration has a non-zero $\vev{A_y}$, 
it will appear through the spurion superfield~$T$ in Eq.(\ref{action_fml2}). 
For instance, if $\arg(\phcl)$ varies with $\cO(1)$ amplitude 
as $y$ goes from $0$ to $\pi r$, it is roughly estimated as 
\be
 m_g = \int\dr y\; \frac{e^{3A}\vev{A_y}}{2\pi r} 
  \sim \kp^2\Lmd^3.  \label{est_mg}
\ee

Here, we have used the kink mass terms to localize the zero-modes 
of the matter fields. 
We can also localize them by introducing the coupling between 
the matters and  $\phi^{\rm h}$, 
just like the fat brane scenario \cite{fat-brane}. 
In this case, however, the SUSY-breaking masses arise through 
the direct coupling to $\phi^{\rm h}$ as discussed in Ref.~\cite{sakamura2}, 
and they become the dominant SUSY-breaking effects 
in the visible sector. 
So we do not consider such a possibility in this paper.

\subsection{Orbifold compactification}
Before ending this section, we will briefly comment on the case that 
the extra dimension is compactified on the orbifold~$S^1/Z_2$. 
The $Z_2$-transformation is $y\to -y$, 
and the fixed-point boundaries are at $y=0$ and $y=\pi r$. 
Under the $Z_2$ transformation, each field is transformed as follows. 
\begin{description}
\item[Scalar field~$\phi$ :]
\be
 \phi(x^m,y) \to \Pi_\phi \bar{\phi}(x^m,-y). 
\ee

\item[Spinor field~$\chi^\alp$ :]
\be
 \chi^\alp(x^m,y) \to \Pi_\chi\bar{\chi}^{\dot{\alp}}(x^m,-y). 
\ee
Note that there is no distinction between the dotted and undotted indices 
in three dimensions. 
So the orbifold identification is consistent on the 3D boundaries. 

\item[Vector field~$B^g_\mu$ :]
\bea
 B^g_m(x^m,y) &\to& \Pi_B B^g_m(x^m,-y), \nonumber\\
 B^g_y(x^m,y) &\to& -\Pi_B B^g_y(x^m,-y). 
\eea
\end{description}

$\Pi_\phi$, $\Pi_\chi$ and $\Pi_B$ are the eigenvalues 
of the $Z_2$-parity for the corresponding fields. 
$\Pi_B$ must be chosen to $+1$ in order for the effective theory 
to have the 3D gauge field.
Note that $B^g_y$ can be eliminated completely in this case. 
As mentioned in Ref.~\cite{sakamura1}, 
we can eliminate $B^g_y$ except its zero-mode 
by using the 4D gauge transformation, 
and the zero-mode is projected out by the orbifold identification. 

From the above transformation properties, 
the orbifold identification explicitly breaks the bulk 4D $\cN=1$ SUSY 
to 3D $\cN=1$ SUSY. 

If we assume the $Z_2$-parity eigenvalues for each field as 
\bea
 \Pi=+1 & : & \cA^{I\neq 1},\;\;\; \chi^{I\neq 1},\;\;\; B^g_\mu \nonumber\\ 
 \Pi=-1 & : & \phi^{I=1},\;\;\; \chi^{I=1},\;\;\; \lmd^g, 
\eea
the non-BPS solution of Ref.~\cite{EMS} is allowed. 
In this case, a a half of the bulk SUSY is explicitly broken 
by the orbifold projection, and the rest of it is broken 
by the non-BPS background.

\section{Summary and discussions} \label{summary}
We have discussed the effects of SUSY breaking mediated by 
the deformation of the space-time geometry 
due to the backreaction of a bulk scalar field. 
Since the superfield formalism is very useful, 
we derived the effective action expressed by 
the superfields and the SUSY breaking terms. 
The main obstacle in this procedure is the fact that we cannot define 
the supersymmetry on the non-BPS background geometry in the ordinary sense. 
We have solved this difficulty by defining the global SUSY 
{\it before} the gauge fixing of the superconformal symmetry. 
Since this SUSY transformation does not preserve 
the gauge fixing conditions, 
the effective theory has SUSY breaking terms. 
Namely, in our method, {\it SUSY is broken by the gauge fixing conditions}. 

Let us summarize the procedure to obtain the effective action. 
We have assumed the existence of the non-BPS solution in the 4D SUGRA model. 
\begin{enumerate}
 \item Define 3D global SUSY transformation that preserves 
  the gravitational background. 

 \item Construct superfields by using the above transformation.  

 \item Rewrite the action on the gravitational background 
  in terms of the superfields. 

 \item Rewrite the components of the compensator superfield 
  in terms of the physical fields by the gauge fixing conditions. 
  \label{put_GF}

 \item Substitute the mode-expanded expressions 
  and perform the $y$-integration, 
  then we can obtain the effective action. 
%
\end{enumerate}
At Step~\ref{put_GF}, we have dropped the fluctuation modes of 
the hidden sector field~$\Phi^{I=1}$ around the background 
because we are interested only in the visible sector. 

Note that all the SUSY breaking terms involve at least one of 
the SUSY-breaking functions~$\fhcl$, $\fG$ and $\vev{A_y}$. 
In Eq.(\ref{Smatter}), SUSY breaking scalar masses and 
scalar trilinear couplings come from 
only the superpotential of the bulk 4D theory, 
but this is due to the assumption that the K\"{a}hler potential 
is minimal. 
Since the compensator multiplet 
does not couple to the superfield strength multiplet, 
no gaugino mass arises from the deformed geometry 
if $\vev{A_y}=0$. 

Eq.(\ref{Smatter}) shows that 
there is no BPS solution with $\vev{A_y}\neq 0$. 
In other words, if we consider the case of $\vev{A_y}\neq 0$, 
the corresponding background is necessarily non-BPS, 
and $\fhcl$ and $\fG$ also do not vanish. 

Scales introduced in the bulk theory are the 4D Planck scale~$\Mp$, 
the mass parameters for the matter fields~$m_i$, 
the characteristic scale of the hidden sector dynamics~$\Lmd$, 
and the compactification scale~$r^{-1}$. 
In terms of these scales, 
the SUSY-breaking scalar masses induced in the visible sector 
have a form of 
\be
 m_S^2=\frac{\Lmd^3m_i}{\Mp^2}\alp(m_i,\Lmd,r), 
 \label{SBscl_est1}
\ee
where $\alp(m_i,\Lmd,r)$ is a dimensionless function expressed by 
the overlap integral of the mode-functions 
and the SUSY-breaking functions~$\fhcl$, $\fG$ and $\vev{A_y}$. 
The gaugino mass is induced by the non-zero $\vev{A_y}$, 
and roughly estimated as 
\be
 m_g\sim \frac{\Lmd^3}{\Mp^2}. \label{est_mg2}
\ee
Here, we have supposed that $\arg(\phcl)$ varies with $\cO(1)$ amplitude 
as $y$ goes from $0$ to $\pi r$. 
These estimations are modified if the hidden sector 
involves more than one mass scales. 

For example, in the case that $m_i\sim\Lmd$ and $\alp=\cO(1)$,
the above expressions mean that 
the gaugino mass is much smaller than the scalar masses if $\Mp\gg\Lmd$. 
However, as seen in Sect.~\ref{scales}, $\alp$ can be exponentially small. 
In such a case, the gaugino mass can dominate the scalar masses. 

If the warp factor is not so large, 
the gaugino mass is also estimated from Eq.(\ref{est_mg}) as 
\be
 m_g\sim\frac{\abs{\Dlt\vtht_A}}{r}, \;\;\;\;\;
 \Dlt\vtht_A\equiv\left\{\begin{array}{ll} \vtht_A(\pi r)-\vtht_A(-\pi r) & 
  \mbox{for $S^1$-compactification} \\
  \vtht_A(\pi r)-\vtht_A(0) & \mbox{for $S^1/Z_2$-compactification} 
  \end{array}\right. 
\ee
where $\vtht_A(y)$ is defined in Eq.(\ref{def_xiA}). 
Comparing this and Eq.(\ref{est_mg2}), we can see that 
$r^{-1}\sim \Lmd^3/\Mp^2\ll\Lmd$ when $\Dlt\vtht_A=\cO(1)$ and 
$\Mp\gg\Lmd$. 
In fact, since $\Lmd^{-1}$ is the characteristic length of 
the nontrivial field configuration, the radius~$r$ is generally 
larger than $\Lmd^{-1}$. 


Since the SUSY breaking effects are suppressed by the Planck scale, 
our scenario can be considered as a kind of the gravity mediation. 
However, there are some points that should be noticed. 
First, from the viewpoint of the effective theory, SUSY breaking cannot be 
regarded as the spontaneous breaking 
because the order parameter of SUSY breaking~$\Lmd_S$ is roughly 
of $\cO(\Lmd)$ and is generally higher 
than the compactification scale~$r^{-1}$, 
which is the cut-off scale of the 3D effective theory. 
Second, the induced SUSY breaking scale in the effective theory 
can be suppressed by the overlap integral of 
the mode functions and the SUSY-breaking functions.  
(See $\alp$ in Eq.(\ref{asymp_alp}).) 
Third, the gaugino mass can be induced by non-zero $\vev{A_y}$ 
without introducing a non-minimal gauge kinetic function. 
This is very similar to the Scherk-Schwarz (SS) SUSY breaking 
in the flat space-time. 
However, this breaking is irrelevant to the $U(1)_R$ twisting 
since $U(1)_R$ is a symmetry {\it after} the gauge fixing 
and independent of $U(1)_A$, which is completely fixed 
by the gauge fixing condition.  
In addition, $\vev{A_y}$ is not an input parameter as in the SS breaking, 
but is determined by the hidden sector dynamics 
and has a nontrivial $y$-dependence. 
Further, the non-zero $\vev{A_y}$ indicates that the background configuration 
is non-BPS as mentioned above. 
Thus, it inevitably leads to SUSY breaking terms that associate with 
$\fhcl$ and $\fG$, to which the SS breaking does not have any resemblances. 


To investigate more phenomenological aspects, 
we should extend our discussion to 5D SUGRA. 
Note that the procedure explained in this paper requires only 
the knowledge of the superconformal formulation of 4D SUGRA
and the 3D superfield formalism.  
The 4D superfield formalism is not necessary. 
Therefore, when we extend our method to 5D SUGRA, 
we do not need to deal with the unknown 5D superfield formalism. 
The 5D superconformal formulation and the well-known 4D $\cN=1$ superfield 
formalism are enough. 
The former has already been studied 
in Ref.~\cite{KO,FO}, and the 5D invariant action is provided 
in Ref.~\cite{KO2}. 
So we can utilize their results. 
Our strategy in this paper can basically be extended to 
5D SUGRA straightforwardly, but there are some subtle points. 
For example, all hypermultiplets are charged under the central charge 
group~$U(1)_Z$ \cite{FO}, which has no corresponding group in the 4D case. 
The $U(1)_Z$ transformation of the hypermultiplets is highly nontrivial, 
and thus some special treatment for the $U(1)_Z$ gauge multiplet 
is necessary in the superfield formalism. 
However, we expect that these subtle points do not cause serious difficulties, 
and can be solved by slight modifications of our method. 
Research along this direction is now in progress. 
Including the brane-localized terms in the discussion 
is also an interesting issue. 
We will discuss these subjects in the subsequent paper.

\vspace{5mm}

\begin{center}
{\bf Acknowledgments}
\end{center}
The author would like to thank Hiroyuki Abe and Keisuke Ohashi 
for useful discussion and advise.  
The author is indebted to Tatsuya Noguchi for a collaboration 
in an early stage.   
This work was supported from the Astrophysical Research Center 
for the Structure and Evolution of the Cosmos (ARCSEC) 
funded by the Korea Science and Engineering Foundation 
and the Korean Ministry of Science.

\appendix
\section{Notations} \label{notations}
\subsection{Superconformal formulation of 4D SUGRA}\label{scSUGRA}
Basically, we follow the notations of Ref.~\cite{WB} 
for the four-dimensional bulk theory. 

Our notations can be obtained from those of Ref.~\cite{KO} 
by rewriting spinors into the 2-component representation 
and replacing each quantity as follows. 
\begin{description}
 \item[Metric:]
\be
 \eta_{ab} \to -\eta_{ab}. \label{replace1}
\ee

\item[Superconformal gauge fields:]
\bea
 \ge{\mu}{a} &\to& \ge{\mu}{a}, \;\;\;
 \psi_\mu^\alp \to \frac{\kp}{2}\psi_\mu^\alp, \;\;\;
 b_\mu \to b_\mu, \;\;\;
 A_\mu \to -\frac{4}{3}A_\mu, \nonumber\\
 \gomg{\mu}{ab} &\to& -\gomg{\mu}{ab}, \;\;\;
 \phi_\mu^\alp \to i\vph_\mu^\alp, \;\;\;
 \gf{\mu}{a} \to -\gf{\mu}{a}. 
\eea

\item[Real general multiplet:]
\bea
 C &\to& C, \;\;\;
 \zeta \to -\zeta, \;\;\;
 \frac{1}{2}\brkt{H+iK} \to i\cH, \nonumber\\
 B_a &\to& -B_a, \;\;\;
 \lmd_\alp \to \lmd_\alp, \;\;\;
 D \to D. 
\eea

\item[Chiral multiplet:]
\be
 {\cal A} \to \cA, \;\;\;
 \chi_\alp \to -\sqrt{2}\chi_\alp, \;\;\;
 \cF \to -\cF. 
 \label{replace4}
\ee
\end{description}

In the following, we will collect some important expressions in our notations. 
\begin{description}
\item[4D superconformal algebra:]
\bea
 \sbk{M_{ab},M_{cd}}\eql i\brkt{\eta_{ad}M_{bc}+\eta_{bc}M_{ad}
 -\eta_{ac}M_{bd}-\eta_{bd}M_{ac}}, \nonumber\\
 \sbk{M_{ab},P_c}\eql i\brkt{\eta_{bc}P_a-\eta_{ac}P_b}, \;\;\;\;
 \sbk{M_{ab},K_c}=i\brkt{\eta_{bc}K_a-\eta_{ac}K_b}, \nonumber\\ 
 \sbk{D,P_a}\eql -iP_a, \;\;\; \sbk{D,K_a}=iK_a, \;\;\;\
 \sbk{P_a,K_b}= 2iM_{ab}-2i\eta_{ab}D, \nonumber\\
 \sbk{A,Q_\alp}\eql Q_\alp, \;\;\;\; \sbk{A,S_\alp}=-S_\alp, \;\;\;\;
  \nonumber\\
 \sbk{P_a,S_\alp} \eql -i\brkt{\sgm_a}_{\alp\dalp}\bar{Q}^{\dalp}, \;\;\;\;
 \sbk{K_a,Q_\alp}= -i\brkt{\sgm_a}_{\alp\dalp}\bar{S}^{\dalp}, \;\;\;\;
 \nonumber\\
 \sbk{M_{ab},Q_\alp}\eql i\brkt{\sgm_{ab}}_\alp^{\;\bt}Q_\bt, \;\;\;\;
 \sbk{M_{ab},S_\alp}=i\brkt{\sgm_{ab}}_\alp^{\;\bt}S_\bt, \nonumber\\
 \sbk{D,Q_\alp}\eql -\frac{i}{2}Q_\alp, \;\;\;\;
 \sbk{D,S_\alp}=\frac{i}{2}S_\alp, \;\;\;\;
 \nonumber\\
 \brc{Q_\alp,\bar{Q}_{\dalp}}\eql 2\brkt{\sgm^a}_{\alp\dalp}P_a, \;\;\;\;
 \brc{S_\alp,\bar{S}_{\dalp}}=2\brkt{\sgm^a}_{\alp\dalp}K_a, \nonumber\\
 \brc{Q_\alp,S_\bt}\eql -2\ep_{\bt\gm}\brkt{\sgm^{ab}}_\alp^{\;\;\gm}M_{ab}
 +2\ep_{\alp\bt}D+3i\ep_{\alp\bt}A, 
 \label{SCalg}
\eea
All the other commutators vanish. 

\item[Transformation law of the gravitational multiplet:]
\bea
 \dQ\ge{\mu}{a} \eql i\kp\brkt{\ep\sgm^a\bar{\psi}_\mu
  +\bar{\ep}\bsgm^a\psi_\mu}-\lmd_D\ge{\mu}{a}, \nonumber\\
 \dQ\psi_\mu^\alp \eql 2\kp^{-1}\cD^{\rm h}_\mu\ep^\alp
  -\frac{1}{2}\lmd_D\psi_\mu^\alp-i\vtht_A\psi_\mu^\alp
  +\brkt{\bar{\eta}\bsgm_\mu}^\alp, \nonumber\\
%
%
 \dQ A_\mu \eql \der_\mu\vtht_A-3\brkt{\ep\vph_\mu+\bar{\ep}\bar{\vph}_\mu}
  +\frac{3\kp}{2}\brkt{\eta\psi_\mu+\bar{\eta}\bar{\psi}_\mu}, 
\eea
where $\cD^{\rm h}$ is a covariant derivative with respective to 
the homogeneous transformations~$M_{ab},D,A$. 
\be
 \cD^{\rm h}_\mu\ep^\alp = \der_\mu\ep^\alp-\frac{1}{2}\gomg{\mu}{ab}
  \brkt{\ep\sgm_{ab}}^\alp+\frac{1}{2}b_\mu\ep^\alp+iA_\mu\ep^\alp. 
\ee
Transformation laws of the other multiplets can be read off 
from Ref.~\cite{KO} by the replacements~(\ref{replace1})-(\ref{replace4}). 

\item[Multiplication laws of chiral multiplets:]\mbox{}\\
From a set of chiral multiplets~$\Sgm^I=[\cA^I,\chi^I_\alp,\cF^I]$, 
we can produce a new chiral multiplet~$W(\Sgm)$ 
where $W$ is an arbitrary function. 
\be
 W(\Sgm) = [W(\cA),W_I\chi^I_\alp,
  -\frac{1}{2}W_{IJ}\chi^I\chi^J+W_I\cF^I]. 
\ee
If a function~$G$ involves both chiral and anti-chiral multiplets, 
it induces a general multiplet~$G(\Sgm,\bar{\Sgm})
=[C^G,\zeta^G_\alp,\cH^G,B^G_a,\lmd^G_\alp,D^G]$ with 
\bea
 C^G \eql G(\cA,\bar{\cA}), \nonumber\\
 \zeta^G_\alp \eql -\sqrt{2}iG_I\chi^I_\alp, \nonumber\\
 \cH^G \eql \frac{i}{2}G_{IJ}\brkt{\chi^I\chi^J}-iG_I\cF^I, \nonumber\\
 B^G_a \eql -iG_I\cD_a\cA^I+iG_{\bar{J}}\cD_a\bar{\cA}^{\bar{J}}
  +G_{I\bar{J}}\brkt{\chi^I\sgm_a\bar{\chi}^{\bar{J}}}, \nonumber\\
 \lmd^G_\alp \eql \sqrt{2}G_{I\bar{J}}\brc{
  \brkt{\sgm^a\bar{\chi}^{\bar{J}}}_\alp\cD_a\cA^I
  +i\bar{\cF}^{\bar{J}}\chi^I_\alp}-\frac{i}{\sqrt{2}}G_{I\bar{J}\bar{K}}
  \brkt{\bar{\chi}^{\bar{J}}\bar{\chi}^{\bar{K}}}\chi^I_\alp, 
 \nonumber\\
 D^G \eql G_{I\bar{J}}\brc{-2\cD^a\bar{\cA}^{\bar{J}}\cD_a\cA^I
  -i\chi^I\sgm^a\cD_a\bar{\chi}^{\bar{J}}-i\bar{\chi}^{\bar{J}}\bar{\sgm}^a
  \cD_a\chi^I+2\bar{\cF}^{\bar{J}}\cF^I} \nonumber\\
 &&+G_{IJ\bar{K}}\brc{
  i\chi^J\sgm^a\bar{\chi}^{\bar{K}}\cD_a\cA^I-\bar{\cF}^{\bar{K}}
  \brkt{\chi^I\chi^J}} \nonumber\\
 &&+G_{I\bar{J}\bar{K}}\brc{
  i\bar{\chi}^{\bar{K}}\bar{\sgm}^a\chi^I\cD_a\bar{\cA}^{\bar{J}}
  -\cF^I\brkt{\bar{\chi}^{\bar{J}}\bar{\chi}^{\bar{K}}}} 
  +\frac{1}{2}G_{IJ\bar{K}\bar{L}}\brkt{\chi^I\chi^J}
  \brkt{\bar{\chi}^{\bar{K}}\bar{\chi}^{\bar{L}}}, 
\eea
where $\cD_\mu$ is the superconformal covariant derivative. 

\item[$F$-term action formula:]\mbox{}\\
For a chiral multiplet~$\Phi^{(w=3)}=[\cA,\chi,\cF]$ with weight~$w=3$ 
($n=2$), 
we can construct the following superconformal invariant action: 
\be
 S_F=\int\dr^4x \sbk{\Phi^{(w=3)}}_F
 \equiv\int\dr^4x e\sbk{\cF+\sqrt{2}i\bar{\psi}_a\bar{\sgm}^a\chi
  -4\bar{\psi}_a\bar{\sgm}^{ab}\bar{\psi}_b\cA+\hc}.  
 \label{F_action_fml}
\ee

\item[$D$-term action formula:]\mbox{}\\
For a real general multiplet~$V=[C,\zeta_\alp,\cH,B_a,\lmd_\alp,D]$ 
with weight~$w=2$, $n=0$, we can construct the following invariant action:
\bea
 S_D \eql \int\dr^4x\sbk{V^{(w=2,n=0)}}_D \nonumber\\
 \defa \int\dr^4x e\left[D-\psi_a\sgm^a\bar{\lmd}+\bar{\psi}_a\bsgm^a\lmd
 -2i\ep^{abcd}\psi_a\sgm_b\bar{\psi}_c
 \brkt{B_d+\psi_d\zeta+\bar{\psi}_d\bar{\zeta}} \right. \nonumber\\
 &&\hspace{20mm}
 +\frac{1}{3}\brc{R(\omg)+4e^{-1}\ep^{abcd}
 \brkt{\psi_a\sgm_d\cDh_b\bar{\psi}_c-\bar{\psi}_a\bar{\sgm}_d\cDh_b\psi_c}}
 C \nonumber\\
 &&\hspace{20mm}\left.
 +\frac{4i}{3}\brkt{\zeta\sgm^{ab}\cDh_a\psi_b
 -\bar{\zeta}\bsgm^{ab}\cDh_a\bar{\psi}_b}\right], 
 \label{D_action_fml}
\eea
where
\bea
 R(\omg)\defa \ge{a}{\mu}\ge{b}{\nu}R_{\mu\nu}^{\;\;\;\;ab}(\omg)
 =2\ge{a}{\mu}\ge{b}{\nu}\der_\mu\gomg{\nu}{ab}
  +\gomg{b}{ac}\omg_{ac}^{\;\;\;\;b}-\gomg{a}{ac}\omg_{bc}^{\;\;\;\;b}, 
 \nonumber\\
 \cDh_\mu\psi_\nu^\alp\defa \der_\mu\psi_\nu^\alp
  -\frac{1}{2}\gomg{\mu}{ab}\brkt{\psi_\nu\sgm_{ab}}^\alp
  +\frac{1}{2}b_\mu\psi_\nu^\alp+iA_\mu\psi_\nu^\alp. 
\eea

\item[Gauge fixing conditions:]\mbox{}\\
The first condition in Eq.(\ref{scGF}) comes from the requirement 
that the Einstein term should be canonically normalized.  
Namely, $C$ in Eq.(\ref{D_action_fml}) should be fixed as 
\be
 C=-\bar{\cA}^{\bar{\Sgm}}\cA^\Sgm\exp\brc{-\frac{\kp^2}{3}
  \sum_{I\neq 0}\bar{\phi}^{\bar{I}}\phi^I}=-\frac{3}{2\kp^2}. 
\ee
This condition does not determine the phase of $\cA^\Sgm$.  
Here, we have chosen it so that $\cA^\Sgm$ is real.  

The second condition comes from the requirement that 
a mixing kinetic term between the gravitino~$\psi_\mu$ 
and the chiral fermions~$\chi^I$ should vanish. 
Namely, the condition is 
\be
\zeta_\alp=-\sqrt{2}i\bar{\cA}^{\bar{\Sgm}}\exp\brc{-\frac{\kp^2}{3}
  \sum_{I\neq 0}\bar{\phi}^{\bar{I}}\phi^I}\cdot
  \brkt{\frac{\kp^2}{3}\cA^\Sgm\sum_{J\neq 0}\bar{\phi}^{\bar{I}}\chi^I_\alp
  -\chi^\Sgm_\alp}=0. 
\ee

Under the above two conditions, the dependence of the action~$S$ on $b_\mu$ 
(the gauge field for the dilatation~$D$) becomes a surface term 
and vanishes. 
\be
 S|_{b_\mu}=-2\int\dr^4 x\; \der_\mu(eb^\mu)=0. 
\ee 
Therefore, we can fix $b_\mu$ to an arbitrary value. 
Here, we have fixed it to zero. 
\end{description}

\subsection{Notations for 3D theory} \label{3Dnotations}
The notations for the 3D theories are as follows. 

We take the space-time metric as 
\be
 \eta_{\udl{m}\udl{n}}=\diag(-1,+1,+1), 
\ee
where $\udl{m},\udl{n}=0,1,3$. 

The 3D $\gm$-matrices, $(\gm_{(3)}^{\udl{m}})_\alpha^{\;\;\beta}$, 
can be written by the Pauli matrices as  
\be
 \gm_{(3)}^{\udl{0}}=\sgm^2, \;\;\;
 \gm_{(3)}^{\udl{1}}=-i\sgm^3, \;\;\;
 \gm_{(3)}^{\udl{3}}=i\sgm^1, 
\ee
and these satisfy the 3D Clifford algebra, 
\be
 \{\gm_{(3)}^{\udl{m}},\gm_{(3)}^{\udl{n}}\}=-2\eta^{\udl{m}\udl{n}}.
\ee

The generators of the Lorentz group~$Spin(1,2)$ are 
\be
 \gm_{(3)}^{\udl{m}\udl{n}}\equiv
  \frac{1}{4}[\gm_{(3)}^{\udl{m}},\gm_{(3)}^{\udl{n}}]. 
\ee

The relations between the 4D $\sgm$-matrices and 
the above $\gm_{(3)}^{\udl{m}}$ are 
\bea
 (\sgm^a)_{\alpha\dot{\beta}} \eql 
 (\gm_{(3)}^{\udl{0}},\gm_{(3)}^{\udl{1}},-1,
  \gm_{(3)}^{\udl{3}})_\alpha^{\;\;\gm}(-\sgm^2)_{\gm\beta}, \nonumber\\
 (\bar{\sgm}^a)^{\dot{\alpha}\beta} \eql 
 (-\sgm^2)^{\alpha\gm}
 (\gm_{(3)}^{\udl{0}},\gm_{(3)}^{\udl{1}},1,
  \gm_{(3)}^{\udl{3}})_\gm^{\;\;\beta}, 
 \label{sgm_gm3}
\eea
\bea
 (\sgm^{\udl{m}\udl{n}})_\alpha^{\;\;\beta} \eql 
 (\gm_{(3)}^{\udl{m}\udl{n}})_\alpha^{\;\;\beta}, \nonumber\\
 (\sgm^{\udl{m}\udl{2}})_\alpha^{\;\;\beta} \eql 
 \frac{1}{2}(\gm_{(3)}^{\udl{m}})_\alpha^{\;\;\beta}. 
\eea
Note that, in three dimensions, there is no discrimination 
between the dotted and undotted indices. 

The spinor indices are raised and lowered by multiplying $\sgm^2$ 
from the left. 
\be
 \psi_\alpha=(\sgm^2)_{\alpha\beta}\psi^\beta, \;\;\;
 \psi^\alpha=(\sgm^2)^{\alpha\beta}\psi_\beta. 
\ee
We take the following convention of the contraction of spinor indices. 
\be
 \psi_1\psi_2\equiv \psi_1^\alpha \psi_{2\,\alpha}
 =(\sgm^2)_{\alpha\beta}\psi_1^\alpha \psi_2^\beta
 =\psi_2\psi_1. 
\ee

\section{Radion superfield} \label{radion_sf}
To identify the radion superfield, we will revive the extra components 
of the Weyl multiplet~$\ge{y}{\udl{2}}$, $\psi_y^\alp$ and $A_y$ 
as fields, while other components are kept to be their background 
values. 
Then, we will consider the transformation laws of them 
under the global SUSY transformation~$\dQ(\ep_0)$ 
defined by Eqs.(\ref{sc_dlt}) and (\ref{gl_prmt}). 
If we decompose $\psi_y^\alp$ as 
\be
 \psi_y^\alp=\frac{e^{i\vtht_0}}{\sqrt{2}}\brkt{\psi_{{\rm R}y}^\alp
  +i\psi_{{\rm I}y}^\alp}, 
\ee
we can easily show that 
\be
 \dQ(\ep_0)\psi_{{\rm I}y}^\alp = 0. 
\ee
So we can put the gauge fixing condition, 
\be
 \psi_{{\rm I}y}^\alp=0. 
\ee
Under this gauge fixing, $\dQ(\ep_0)$ is closed among the extra components 
of the Weyl multiplet, and we can construct a following superfield. 
\be
 T\equiv e^{\dQ(\tht)}\ge{y}{\udl{2}}
 =\ge{y}{\udl{2}}-\kp e^{\frac{A}{2}}\tht\psi_{{\rm R}2}
  +\frac{e^A}{2}\tht^2(2A_y). 
\ee
This corresponds to the radion superfield. 
However, when a bulk scalar field is responsible for the radius stabilization, 
its fluctuation is entangled with the fluctuation of $\ge{y}{\udl{2}}$. 
Thus, we have to diagonalize the coupled system to obtain 
the physical radion \cite{kofman}. 

\end{document}